# Electromagnetically Induced Entanglement

Xihua Yang[1], and Min Xiao[2,3]

[1]*Department of Physics, Shanghai University, Shanghai 200444, China*

[2]*National Laboratory of Solid State Microstructures and School of Physics, Nanjing University, Nanjing 210093, China*

[3]*Department of Physics, University of Arkansas, Fayetteville, Arkansas 72701, USA*

(Dated: May 5, 2015)

We present a novel quantum phenomenon named electromagnetically induced entanglement in the conventional Λ-type three-level atomic system driven by a strong pump field and a relatively weak probe field. Nearly perfect entanglement between the pump and probe fields can be achieved with a low coherence decay rate between the two lower levels, high pump-field intensity, and large optical depth of the atomic ensemble. The physical origin is quantum coherence between the lower doublet produced by the pump and probe fields, similar to the well-known electromagnetically induced transparency. This method would greatly facilitate the generation of nondegenerate narrow-band continuous-variable entanglement between bright light beams by using only coherent laser fields, and may find potential and broad applications in realistic quantum information processing.

The strong interactions between material media and laser fields may lead to many special effects, such as coherent population trapping (CPT) [1], electromagnetically induced transparency (EIT) [2], rapid adiabatic population transfer [3], enhancement of nonlinear optical processes [4], and quantum correlation and entanglement [5-11]. The common physical mechanism underlying these peculiar effects is the laser-induced quantum coherence established in the media. Though the above-mentioned effects have been extensively examined in the Λ- or multi-Λ-type atomic systems driven by two or more laser fields, where at least one of the fields is generally treated classically, only a few studies have been performed on the quantum correlated properties of the pump and probe fields in the simplest and widely-employed Λ-type three-level atomic system [12-14]. The phase and amplitude correlations between the two laser fields in the Λ-type three-level atomic system have been investigated in Refs. [12, 13]; it was pointed out by Nussenzveig *et al.* [14] that there exist intrinsic quantum correlation and entanglement between the pump and probe fields in this simple EIT scheme. However, in Ref. [14], they used the interaction between two fields and atoms in a ring cavity to approximately simulate the interaction of atoms with two propagating fields.

In this paper, we directly consider the interaction of an ensemble of atoms with the propagating pump and probe fields on the basis of Heisenberg-Langevin equations, and demonstrate the novel effect, named electromagnetically induced entanglement (EIE), in the Λ-type three-level atomic system. It is shown that, in the traditional EIT regime, nearly perfectly correlated and entangled pump and probe fields can be achieved at high pump-field intensity, high optical depth of the atomic medium, and a relatively low coherence decay rate between the two lower levels (as compared to the spontaneous decay rate of the excited state). This method provides an efficient and convenient way to generate bright nondegenerate narrow-band entangled fields by using only coherent light fields, which may hold great promise for practical applications in quantum communication and quantum networks.

The considered Λ-type three-level atomic system driven by a strong pump field

($E_c$) and a relatively weak probe field ($E_p$), as shown in Fig. 1a, is based on the $D_1$ transitions in $^{85}$Rb atom used in the experiment [15]. Levels $|1\rangle$, $|2\rangle$, and $|3\rangle$ correspond, respectively, to the ground-state hyperfine levels $5S_{1/2}$ (F=2), $5S_{1/2}$ (F=3) and the excited state $5P_{1/2}$ in $D_1$ line of $^{85}$Rb atom. The coherent probe field with frequency $\omega_p$ and pump field with frequency $\omega_c$ couple transitions $|1\rangle$ to $|3\rangle$ and $|2\rangle$ to $|3\rangle$ with the frequency detunings given by $\Delta_1 = \omega_p-\omega_{31}$ and $\Delta_2 = \omega_c-\omega_{32}$, respectively, where $\omega_{ij}$ (i ≠ j) is the atomic resonant frequency between levels i and j. We denote $\gamma_1$ and $\gamma_2$ as the population decay rates from level 3 to levels 1 and 2, respectively, and $\gamma_{12}$ as the dephasing rate between levels 1 and 2 due to atomic collisions and/or finite interaction time between atoms and light fields. In what follows, we take into account the quantum features of both the pump and probe fields by using the Heisenberg-Langevin method, and show how EIE in this Λ-type three-level atomic system can be established via the atomic coherence between the lower doublet created by the two fields.

The quantum operators of the probe and pump fields are denoted as $a_1(z,t)$ and $a_2(z,t)$, and the collective atomic operators as $\sigma_{ij}(z,t)$ (i, j=1, 2, and 3), respectively. The interaction Hamiltonian in the rotating-wave approximation has the form [12, 16, 17]

$$\hat{V} = -\frac{\hbar N}{L}\int_0^L dz \left(\Delta_1 \sigma_{33}(z,t) + (\Delta_1 - \Delta_2)\sigma_{22}(z,t) + g_1 a_1(z,t)\sigma_{31}(z,t) + g_2 a_2(z,t)\sigma_{32}(z,t) + H.c.\right)$$

(1)

where $g_{1(2)} = \mu_{13(23)} \cdot \varepsilon_{1(2)}/\hbar$ is the atom-field coupling constant with $\mu_{13(23)}$ being the dipole moment for the 1-3 (2-3) transition and $\varepsilon_{1(2)} = \sqrt{\hbar\omega_{1(2)}/2\epsilon_0 V}$ being the electric field of a single probe (pump) photon; $\epsilon_0$ is the free space permittivity and V is the interaction volume with length L and beam radius r; and N is the total number of atoms in the interaction volume. The Heisenberg-Langevin equations for describing the evolutions of the atomic operators can be written as

$$\dot{\sigma}_{11}(z,t) = \gamma_1 \sigma_{33} + ig_1 a_1^+(z,t)\sigma_{13} - ig_1 a_1(z,t)\sigma_{31} + F_{11}(z,t), \qquad (2)$$

$$\dot{\sigma}_{22}(z,t) = \gamma_2 \sigma_{33} + ig_2 a_2^+(z,t)\sigma_{23} - ig_2 a_2(z,t)\sigma_{32} + F_{22}(z,t), \qquad (3)$$

$$\dot{\sigma}_{12}(z,t) = -(\gamma_{12} - i(\Delta_1 - \Delta_2))\sigma_{12} - ig_1 a_1(z,t)\sigma_{23}^+ + ig_2 a_2^+(z,t)\sigma_{13} + F_{12}(z,t), \qquad (4)$$

$$\dot{\sigma}_{13}(z,t) = -[\gamma_{13} - i\Delta_1]\sigma_{13} + ig_1 a_1(z,t)(\sigma_{11} - \sigma_{33}) + ig_2 a_2 \sigma_{12} + F_{13}(z,t), \qquad (5)$$

$$\dot{\sigma}_{23}(z,t) = -(\gamma_{23} - i\Delta_2)\sigma_{23} + ig_2 a_2(z,t)(\sigma_{22} - \sigma_{33}) + ig_1 a_1 \sigma_{12}^+ + F_{23}(z,t), \qquad (6)$$

where $\gamma_{13} = \gamma_{23} = \frac{\gamma_1 + \gamma_2}{2}$, and $F_{ij}(z,t)$ are the collective atomic $\delta$-correlated Langevin noise operators. The noise correlation functions for the noise operators $F_{ij}(z,t)$ can be expressed as $\langle F_{ij}(z,t) F_{i'j'}(z',t') \rangle = \frac{L}{N} \mathcal{D}_{ij,i'j'}(z,t)\delta(z-z')\delta(t-t')$, where $\mathcal{D}_{ij,i'j'}(z,t)$ is the Langevin diffusion coefficient and can be calculated by using the generalized dissipation-fluctuation theorem [12]. Under the assumption that the uniformly-distributed pencil-shaped atomic sample is optically thin in the transverse direction, the evolutions of the annihilation operators $a_1$ and $a_2$ can be described by the coupled propagation equations,

$$(\frac{\partial}{\partial t} + c\frac{\partial}{\partial z})a_1(z,t) = ig_1 N \sigma_{13}, \qquad (7)$$

$$(\frac{\partial}{\partial t} + c\frac{\partial}{\partial z})a_2(z,t) = ig_2 N \sigma_{23}. \qquad (8)$$

We use the similar analysis as in Ref. [18] by writing each atomic or field operator as the sum of its mean value and a quantum fluctuation term (i.e., $\sigma_{ij} = \langle \sigma_{ij} \rangle + \delta\sigma_{ij}$ and $a_i = \langle a_i \rangle + \delta a_i$) to treat the interaction between the atoms and the fields. We assume that $\gamma_1, \gamma_2 \gg \gamma_{12}$ and $g_2 \langle a_2 \rangle$ (corresponding to the Rabi frequency of the pump field in semiclassical treatment) is much larger than $\sqrt{\gamma_{12}\gamma_{13}}$ (i.e., in the EIT regime), so the depletions of the pump and probe fields can be safely neglected. Under these conditions, the steady-state mean values of the atomic operators can be obtained by setting the time derivatives equal to zero and neglecting the noise operators. To get the quantum fluctuations of the atomic operators, we substitute the mean values of the atomic and field operators into the Fourier-transformed Heisenberg-Langevin equations for the quantum fluctuations of the atomic operators. By substituting $\delta\sigma_{13}(z,\omega)$ and $\delta\sigma_{23}(z,\omega)$ and their Hermitian conjugates into the Fourier-transformed coupled propagation equations of Eqs. (7) and (8) and their conjugates, the quantum fluctuations of the output field operators $\delta a_1(L,\omega)$ ($\delta a_1^+(L,\omega)$) and $\delta a_2(L,\omega)$ ($\delta a_2^+(L,\omega)$) with respect to the Fourier frequency $\omega$ can be attained, which are linear combinations of the quantum fluctuations of the input field operators $\delta a_1(0,\omega)$ ($\delta a_1^+(0,\omega)$) and $\delta a_2(0,\omega)$ ($\delta a_2^+(0,\omega)$) and Langevin noise terms. To test the entanglement feature between the output pump and probe fields, we use the entanglement criterion $V_{12}=(\delta u)^2+(\delta v)^2<4$, as proposed in Ref. [19], where $\delta u=\delta x_1+\delta x_2$ and $\delta v=\delta p_1-\delta p_2$ with $\delta x_i=(\delta a_i+\delta a_i^+)$ and $\delta p_i=-i(\delta a_i-\delta a_i^+)$ being the amplitude and phase quadrature fluctuation components of the quantum field operator $a_i$. Satisfying the above inequality is a sufficient demonstration for the generation of bipartite entanglement, and the smaller the correlation $V_{12}$ is, the stronger the degree of the bipartite entanglement becomes. We assume the probe and pump fields to be initially in the coherent states $|\alpha_1\rangle$ and $|\alpha_2\rangle$, so the mean values of the field amplitudes are equal to $\langle\alpha_1\rangle$ and $\langle\alpha_2\rangle$, respectively. In the following, the relevant parameters are scaled with m and MHz, or m$^{-1}$ and MHz$^{-1}$, and set according to the realistic experimental conditions used in Ref. [15] with the atomic density $n=1\times10^{19}$, $r=2\times10^{-4}$, $L=0.06$, and $\gamma_1=\gamma_2=3$.

Figure 2a gives the main result of this study, where the evolution of the correlation $V_{12}$ at zero Fourier frequency as a function of the pump-field intensity $|\alpha_2|^2$ is depicted. Obviously, when the intensity of the pump field is very weak, $V_{12}$ is much larger than 4, so no entanglement exists between the probe and pump fields. With the increase of pump field intensity, $V_{12}$ decreases rapidly and becomes less than 4, which sufficiently demonstrates the generation of genuine bipartite entanglement between the two fields. When the pump field intensity is strong enough so that $g_2\langle\alpha_2\rangle \gg \sqrt{\gamma_{12}\gamma_{13}}$, $V_{12}$ becomes nearly equal to zero, which indicates that perfectly quantum-correlated and entangled probe and pump fields can be obtained. For comparison, the absorption coefficient $\alpha$ of the probe field at zero Fourier frequency ($\alpha \propto (\mathrm{Im}\langle\sigma_{31}\rangle)\gamma_{13}/g_1\langle\alpha_1\rangle$ with $\mathrm{Im}\langle\sigma_{31}\rangle$ being the imaginary part of $\langle\sigma_{31}\rangle$) as a function of the pump field intensity is shown in Fig. 2b. It is clear that EIT is exhibited with the increase of pump field intensity. Note that in the above calculations, in order to keep the mean values of atomic operators and the intensity absorption rates of the pump and probe fields nearly stable so as to compare the degrees of entanglement under different pump field intensities, according to the results in Ref. [12], the ratios of $\alpha_2/\alpha_1$, $\gamma_{12}/\alpha_2$, and $n/\alpha_2$ should be kept fixed under the conditions of $\gamma_1, \gamma_2 \gg \gamma_{12}$ and $g_2\langle\alpha_2\rangle \gg \sqrt{\gamma_{12}\gamma_{13}}$, and in our calculations we set the $\alpha_1 = \alpha_2/20$, $n = n_0\alpha_1$, and $\gamma_{12} = 0.1\alpha_1$.

The optical depth of the atomic ensemble plays a key role in the generation of entanglement, as seen from the dependence of the correlation $V_{12}$ at zero Fourier frequency on the atomic density n shown in Fig. 3. Clearly, $V_{12}$, with the initial value of 4, decreases rapidly with the increase of the atomic density, indicating the established bipartite entanglement between the pump and probe fields. When $n_0$ is about larger than $7\times10^{18}/m^3$, $V_{12}$ becomes nearly zero, and the two fields get perfectly entangled. Similar evolution of $V_{12}$ with respect to the interaction length L is also observed. This indicates that the coherent pump and probe beams evolve from initially uncorrelated into nearly perfectly entangled beams as they propagate through the dense atomic medium, accompanying the building up of the atomic coherence between the two lower levels 1 and 2. The requirement of using large optical depth has also been extensively examined for generating entanglement between the Stokes and anti-Stokes fields in the double- and multiple-Λ-type atomic systems [5-11].

The physical mechanism underlying EIE resides in the quantum coherence established between the two lower levels, similar to EIT. This can be seen clearly from Eqs. (2-8). Due to the atomic coherence $\sigma_{12}$ self-produced by the pump and probe fields, the quantum fluctuation of each output field operator comes from a linear combination of the input field operators $\delta a_1(0,\omega)$, $\delta a_1^+(0,\omega)$, $\delta a_2(0,\omega)$, and $\delta a_2^+(0,\omega)$, as well as the Langevin noise terms; therefore, the fluctuations of the pump and probe fields are coupled together and subsequent correlation and entanglement between the two fields can be established as they propagate through the atomic medium. If there were no atomic coherence $\sigma_{12}$, there wouldn't be mutual coupling between the probe and pump fields, and therefore no entanglement would be established between the two fields.

The generated bipartite entanglement between the probe and pump fields can also be well understood in terms of the quantum interference between two $|1\rangle$-$|3\rangle$ transition pathways as shown in Fig. 1b, which is the physical origin of EIT as well. One pathway for the $|1\rangle$-$|3\rangle$ transition is induced by absorbing one probe photon, and another one is induced by absorbing one probe photon and one pump photon and emitting one pump photon. The destructive quantum interference between these two $|1\rangle$-$|3\rangle$ transition pathways results in EIT [2]. As a matter of fact, the second pathway for the $|1\rangle$-$|3\rangle$ transition can be regarded as the result of absorbing one photon from the four-wave-mixing (FWM) field produced by the pump and probe fields; consequently, the interaction between the two $|1\rangle$-$|3\rangle$ transition processes can be equivalently thought to be a closed-loop light-atom interaction of a FWM process. As the photon of the FWM field is identical to the probe field and every generated FWM photon is always accompanied by the depletion of one probe photon and one pump photon, as well as the generation of one pump photon, after this closed-loop light-atom interaction, there are no net losses of energy from both the pump and probe fields, and subsequently both EIT and EIE can be attained in the system. In fact, FWM has already been identified as an efficient process to produce entanglement, and the generations of entangled Stokes and anti-Stokes photons by using FWM processes have been extensively studied in the double- and multi-Λ-type atomic systems [5-11]. As the pump field intensity and the optical depth of the atomic ensemble are increased, the FWM process is strengthened, and therefore the degree of entanglement between the pump and probe fields would be greatly enhanced.

As is well known, EIT critically depends on the dephasing rate $\gamma_{12}$ between the two lower levels, and would deteriorate with the increase of $\gamma_{12}$. This is not the case for EIE under the realistic experimental conditions of $\gamma_1, \gamma_2 \gg \gamma_{12}$ and $g_2\langle\alpha_2\rangle \gg \sqrt{\gamma_{12}\gamma_{13}}$. Figure 4 displays the dependence of the correlation $V_{12}$ at zero Fourier frequency on $\gamma_{12}$. It can be seen that when there is no dephasing between the two lower levels, V equals 4, and no entanglement would exist between the pump and probe fields. With the increase of $\gamma_{12}$, the bipartite entanglement is strengthened dramatically, and nearly perfect entangled pump-probe fields can be achieved. This is due to the fact that when the dephasing $\gamma_{12}$ equals zero, not only $\langle\sigma_{33}\rangle$ but also $\langle\sigma_{13}\rangle$ and $\langle\sigma_{23}\rangle$ will become zero, that is, the atoms will always stay in the dark state and the medium is completely transparent to both the pump and probe fields, so no entanglement can be established between the two fields. Under conditions of $\gamma_1, \gamma_2 \gg \gamma_{12}$, $g_2\langle\alpha_2\rangle \gg \sqrt{\gamma_{12}\gamma_{13}}$ and $\alpha_2 \gg \alpha_1$, as $\langle\sigma_{13}\rangle$ and $\langle\sigma_{23}\rangle$ increase linearly with $\gamma_{12}$ [12], so the probability to leave the dark state will increase with increasing $\gamma_{12}$ and the bipartite entanglement would be strengthened. However, when the dephasing rate $\gamma_{12}$ becomes comparable to or larger than the decay rate $\gamma_1(\gamma_2)$ of the excited state, the dissipation process becomes dominant, and the entanglement would be weakened and eventually disappear (Note that when the dephasing rate $\gamma_{12}$ becomes relatively large, the approximation of neglecting the depletions of the pump and probe fields would break down). Similar behavior has been observed for generating pump-probe intensity correlation [13] and nonclassical (e.g., squeezed) states of light [20] with respect to the dephasing rate $\gamma_{12}$ in the Λ-type CPT configuration.

It is interesting to emphasize that entanglement between two weak light fields (even in the single-photon level) has been predicted in a coherently-prepared atomic medium by using two relatively strong driving laser fields with nearly equal intensity [21], where the two driving fields for creating the atomic coherence are generally treated classically. However, here we show that the two laser fields for pre-establishing atomic coherence in EIT regime are intrinsically entangled themselves under certain conditions. The present result, together with our previous one [11], clearly indicate that arbitrary number of entangled fields including the strong pump and probe fields traditionally-treated classically for preparing atomic coherence can be achieved via multiple FWM processes in the multi-Λ-type atomic systems.

In conclusion, we have demonstrated electromagnetically induced entanglement in the conventional Λ-type three-level atomic system driven by a strong pump field and a relatively weak probe field, which also holds for the ladder- or V-type three-level system. Moreover, this scenario is quite general and may be applied to other systems, such as in solid-state media. The underlying physical mechanism is the quantum coherence between the two lower levels self-produced by the pump and probe fields, similar to the well-known electromagnetically induced transparency. This technique provides a convenient and efficient way to generate nondegenerate narrow-band continuous-variable entanglement between two bright light beams with only coherent laser fields, which may bring great facility in realistic quantum information processing protocols.

## ACKNOWLEDGEMENTS

This work is supported by NBRPC (Nos. 2012CB921804), National Natural Science Foundation of China (Nos. 11274225 and 11321063), Key Basic Research Program of Shanghai Municipal Science and Technology Commission (No. 14JC1402100), and

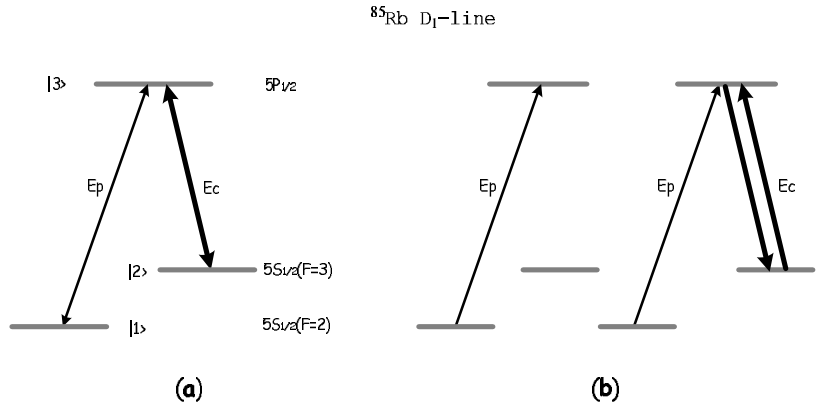

FIG. 1. (**a**) The Λ-type three-level system of the $D_1$ transitions in $^{85}$Rb atom coupled by a strong pump field ($E_c$) and a relatively weak probe field ($E_p$). (**b**) The equivalent configuration of (**a**) for the light-atom interaction resulting in EIT and EIE.

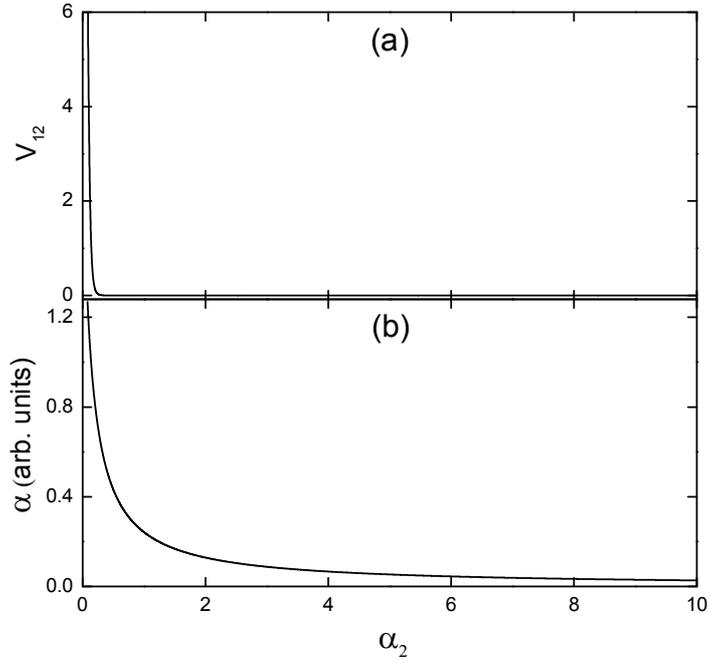

FIG. 2. (a) The evolution of the correlation $V_{12}$ at zero Fourier frequency as a function of the pump field intensity $|\alpha_2|^2$ with $\Delta_1=\Delta_2=0$, $L=0.06$, $r=2.0\times10^{-4}$, $\gamma_1=\gamma_2=3$, $\alpha_1=\alpha_2/20$, $n=1\times10^{19}\alpha_1$, and $\gamma_{12}=0.1\alpha_1$ in corresponding units of m and MHz, or m$^{-1}$ and MHz$^{-1}$. (b) The corresponding absorption coefficient $\alpha$ of the probe field exhibiting EIT.

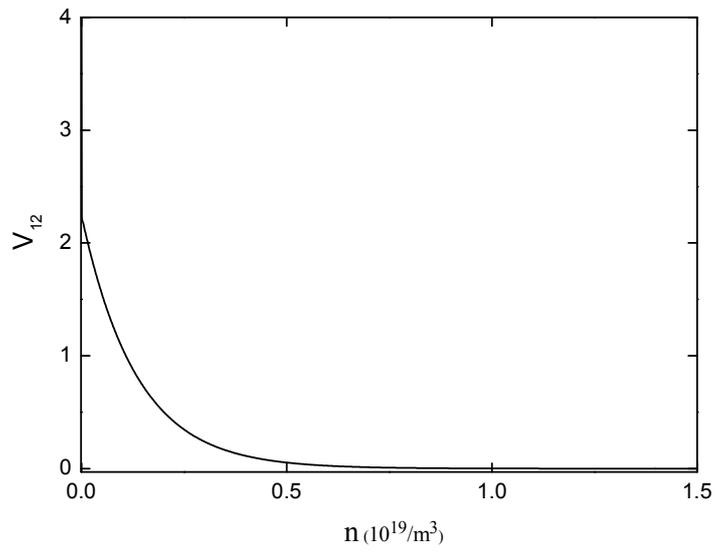

FIG. 3. The dependence of the correlation $V_{12}$ at zero Fourier frequency on the atomic density n with $\alpha_2 = 20\alpha_1 = 20$, $\gamma_{12}=0.1$, and the other parameters are the same as those in Fig. 2.

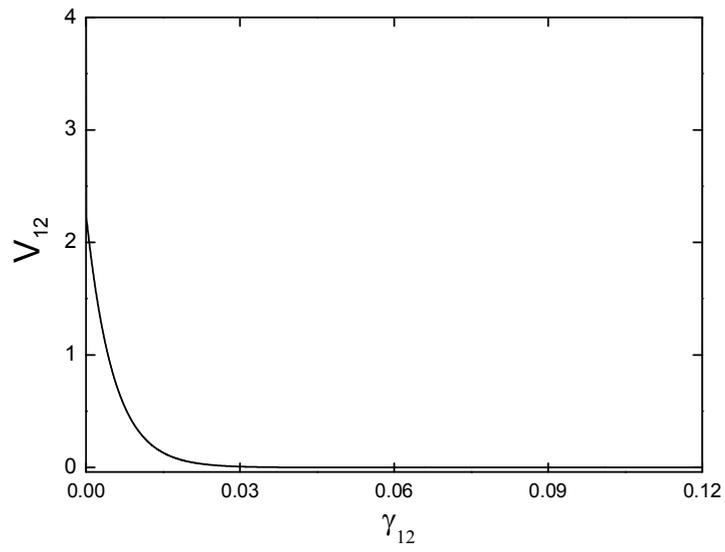

FIG. 4. The dependence of the correlation $V_{12}$ at zero Fourier frequency on the dephasing rate $\gamma_{12}$ with $n=1\times10^{19}$, $\alpha_2 = 20\alpha_1 = 20$, and the other parameters are the same as those in Fig. 2.